\documentclass[twocolumn,showpacs,prl]{revtex4}

\usepackage{graphicx}
\usepackage{dcolumn}
\usepackage{bm}

\begin{document}

\setlength{\baselineskip}{0.4cm}
\addtolength{\topmargin}{1.5cm}

\title{On the Interpretation of ``off the edge'' Avalanches}

\author{O. Ramos$^{1,2}$, A. J. Batista-Leyva$^{1,2,3}$, and E. Altshuler$^{1,2}$}

\affiliation{
$^1$``Henri Poincar\'e" Group of Complex Systems, Physics Faculty, University of Havana, 10400 Havana, Cuba\\
$^2$Superconductivity Laboratory, IMRE-Physics Faculty, University of Havana, 10400 Havana, Cuba \\
$^3$Department of General Physics and Mathematics, InSTEC, P.O.B.
6163 Havana, Cuba}

\date{\today}

\begin{abstract}

We establish both experimentally and theoretically the relation
between {\it off the edge} and {\it internal} avalanches in a
sandpile model, a central issue in the interpretation of most
experiments in these systems. In BTW simulations and also in the
experiments the size distributions of internal avalanches show power
laws and critical exponents related with the dimension of the
system. We show that, in a SOC scenario, the distributions of
{\it off the edge} avalanches do not show power laws but follow
scaling relations with critical exponents different from their analogous
for the internal avalanche distributions.

\end{abstract}


\maketitle

Since Bak, Tang and Wiesenfeld (BTW) developed in 1987 the ideas
of self-organized criticality (SOC)\cite{Bak et al-1987,Bak et
al-1988}, a great amount of research in phenomena as diverse as
earthquakes, superconducting vortex dynamics, stock markets, and
ecology \cite{Bak-1996,Jensen-1998,Altshuler-2000,Altshuler-2004}
has been carried out. A sandpile illustrates this concept: the
slow addition of grains onto a flat surface provokes the growth of
a pile with slopes around a critical angle adjusted through an
avalanche mechanism. According to SOC, the avalanches should not
show any characteristic size or frequency, and the distributions
of avalanche sizes and durations are robust relative to variations
of external parameters; i.e., the system self-organizes. The
result is that the pile will show robust power law distributions
of avalanche size and duration, ``$1/f$" power spectra, and
finite-size scaling of the distribution of {\it internal}
avalanches, measured as the movements of the grains within the
totality of the system.

Although several experiments have been carried out trying to find
the critical behavior predicted from the BTW model in real piles,
the results have not shown a clear agreement. This is due, in
part, to the fact that in most experiments only avalanches that
involve grains abandoning the system (the so-called {\it off the
edge} avalanches) can be measured. Held and co-workers \cite{Held
et al-1990}, using particles between $1.0 - 1.3$ mm, 
reported SOC characteristics when the base of the pile was small
enough, but when it increased, quasiperiodic large avalanches
appeared. This was later corroborated by Rosendahl {\it et
al.}\cite{Rosendahl et al-1993, Rosendahl et al-1994}. With a
similar setup, but using 3 mm diameter spheres, Grumbacher {\it et
al.}\cite{Grumbacher et al-1993} and Costello {\it et
al.}\cite{Costello et al-2003} did not find any lost of SOC
behavior when the relation {\it base diameter/particle size} of
their piles exceeded the value predicted in \cite{Held et al-1990,
Rosendahl et al-1993, Rosendahl et al-1994}. Their piles were also
quite sensitive to the drop height. In ``1D" piles, Frette {\it et
al.}\cite{Frette et al-1996}, using rice with grains of different
shapes, concluded that SOC behavior was attained only for grains
with relative high aspect ratio, because they were able to
decrease inertial effects, not considered in the theory. In
contrast with previous experiments, they were able to measure not
only {\it off the edge} avalanches, but those along the {\it
surface} of the pile. Also with elongated rice grains, but in a 3D
geometry, Aegerter {\it et al.}\cite{Aegerter et al-2003} found scaling
relations measuring the avalanches as variations in the surface of
the system. In our previous work on {\it off the edge}
avalanches in ``1D" piles \cite{Altshuler et al-2001}, the
inertial effects of the 4 mm diameter spheres used were avoided by
introducing strong disorder, reached thanks to a base with beads
glued with random gaps between each other. We claimed that these
piles show SOC behavior, based on the good critical size scaling
of the resulting avalanche distributions and on the fact that the
``active zone" practically involved the whole system. As we have
seen, none of the experiments have measured the avalanches in the
whole system, but only in a portion of it. In most of them, only
the {\it off the edge} avalanches were considered.
\begin{figure}[b]
\vspace{-0.3cm}
\includegraphics[height=1.224in, width=3in]{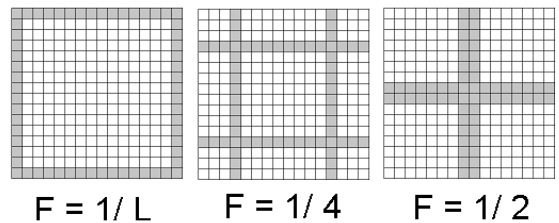}
\vspace{-0.3cm}
\caption{\label{fig:f1} Sites in which avalanches are measured for
different values of $F$ in a $L=16$ grid.}
\end{figure}

In the first part of this paper we measure the avalanches at
several portions of the grid in a two-dimensional BTW computer
simulation, demonstrating that power laws should not be expected
in the size distributions of {\it off the edge} avalanches. After
the avalanche distributions from the simulations are fitted using
an appropriate ansatz, we use the resulting equations to predict
analytically the critical exponents of the corresponding finite
size scalings. The second part reports experiments performed in a
setup similar to that used in \cite{Altshuler et al-2001} where
the position of the center of all beads are obtained after each addition. As a
result, both the internal and {\it off the edge} avalanches are
measured, showing a non power law for the distributions of the
latter, and a power law with the same critical exponents expected
from the 2-D BTW model for the size distributions of internal
avalanches.

We perform a BTW computer simulation on a squared $L\times L$
lattice, following the same rules as in reference \cite{Bak et
al-1987}. The avalanches (defined as the number of sites involved
in toppling events) were measured in the whole grid (internal
avalanches) and also in the rows and columns $LF-1$ sites away
from the borders, where $F$ is a fraction of $L$ (see Fig. 1).
Then, the avalanche distribution for $F=1/L$ coincides with the
{\it off the edge} one. The size distributions for internal and
{\it off the edge} avalanches are shown in Fig. 2. A cursory
inspection of it gives the certainty that the distributions of
{\it off the edge} avalanches are not power laws, while the
internal avalanches exhibit a clear power law behavior over a few
decades. However, internal avalanches and those corresponding to
different values of $F$ are just manifestations of the same
dynamics, so we propose the following ansatz to fit all of their
size distributions
\begin{equation}
  \label{eq:eq1}
  p(s)=p_{0}~s^{-1}\exp(-e1(s/N)^{e2}),
\end{equation}
\noindent where $N$ is the total number of sites considered for
the avalanche measurements and $e2=(L/\sqrt{2} R)^{2}$, where
$R(L)$ is the average distance from the center of the grid to all
the sites of the zone of measurement \cite{next}. $e1$ is constant
for internal avalanches and $e1=A~F^{\gamma}/L^{\xi}$ for
avalanches in the $F$-dependent sites. The fits to the 2-D BTW
simulations presented in Figs. 2 and 3 were reached with $e1=8\pm1$ for
the internal avalanches and $A=4.0\pm0.1$, $\gamma=-0.81\pm0.01$
and $\xi=0.28\pm0.01$ for those in the $F$-dependent sites.

\begin{figure}[b]
\vspace{-0.2cm}
\includegraphics[height=2.4in, width=3.3in]{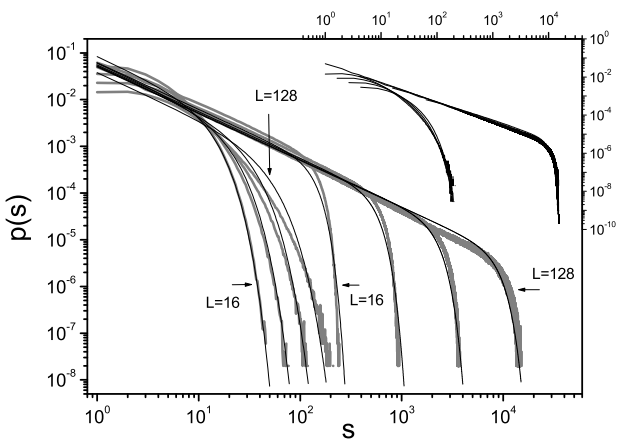}
\vspace{-0.5cm} \caption{\label{fig:f2} Avalanche size
distributions normalized to the total number of avalanches for the
2-D BTW model after $5\cdot10^{7}$ additions, for $L=16,~32,~64$
and 128. The four curves at the right correspond to internal
avalanches, while the four curves at the left correspond to {\it
off the edge} avalanches. Black solid lines fit Eq. (1) to
the data.The scalings based in the ansatz
$p(s,L)L^{\beta}=f(sL^{-\nu})$ are shown in the inset. The
critical exponents appear in Table 1.}
\end{figure}

\begin{figure}[t]
\includegraphics[height=3.65in, width=3.4in]{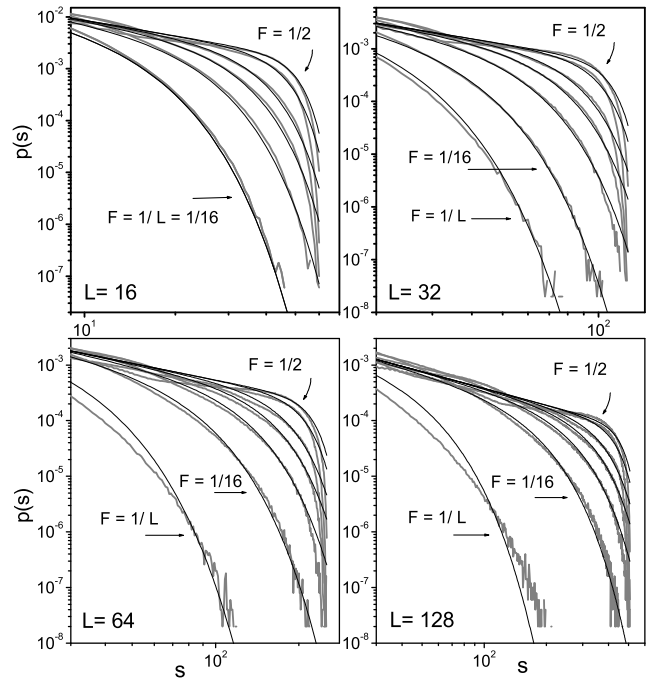}
\vspace{-0.5cm} \caption{\label{fig:f3} Avalanche size
distributions for the $F$-dependent sites (normalized to the total
number of avalanches) for $F$ equal to
$1/2,~6/16,~1/4,~3/16,~1/8,~1/16$ and $1/L$. Black solid lines fit
Eq. (1) to the data.}
\vspace{-0.3cm}
\end{figure}

Having reached the equations that fit the tails of all the
distributions, we can obtain the values of the critical exponents
for their finite size scalings analytically. If we consider a
generic slope $-\alpha$ and we write $N$ as $kL^{d}$, where $d$ is
the dimension of the measurement area, we get
\begin{equation}
  \label{eq:eq2}
  p(s)=p_{0}~s^{-\alpha}\exp(-e1(s/kL^{d})^{e2})
\end{equation}
When the scaling relation $p(s,L)L^{\beta}=f(sL^{-\nu})$ is
applied to a family of curves (in a log-log plot) they move
following a straight line of slope $-\alpha=-\beta/\nu$; which
suggests that if they collapse under scaling relations, it 
is due to the fact that there is an overall power law mechanism
related with them; therefore,
the avalanche distributions of $F$-dependent sites would
have the same critical behavior as the internal ones. In order to
analyze that, let us define $P(S)=p(s)L^{\beta}$ and
$S=sL^{-\nu}$, then
\begin{equation}
  \label{eq:eq3}
  P(S)=p_{0}~S^{-\alpha}L^{\beta-\alpha\nu}\exp(-e1(SL^{\nu-d}/k)^{e2})
\end{equation}
~For internal avalanches $d=2$, $k=1$, and $e1$ is constant; then we have $\nu=d=2$
 and $\beta=\alpha\nu$ . As the slope is $-\alpha=-1$, we get $\beta=2$.
 For the other distributions $p_{0}\sim L^{\eta}$, where
 $\eta=-0.37\pm0.01$. For constant $F$ distributions $d=1$ and $k=4$, then
\begin{equation}
  \label{eq:eq4}
  P(S)\sim S^{-\alpha}L^{\beta-\alpha\nu+\eta}\exp(-AF^{\gamma}(S/4)^{e2}L^{-\xi+e2(\nu-d)})
\end{equation}
\noindent so $\nu=1+\xi/e2$ and $\beta=\alpha\nu-\eta$. We can
consider $\alpha=1$, and calculate the corresponding values for
$\beta$; or consider $\beta=1.45$ (obtained from the values of
$e2$ and $\alpha$ for the internal avalanche distributions), and
calculate the $\alpha$ values. All the critical exponent are shown
in Table 1. For {\it off the edge} avalanches $d=1$, $k=4$ and
$F=1/L$, then
\begin{equation}
  \label{eq:eq5}
  P(S)\sim S^{-\alpha}L^{\beta-\alpha\nu+\eta}\exp(-A(S/4)^{e2}L^{-\gamma-\xi+e2(\nu-d)})
\end{equation}
\noindent thus, $\nu=1+(\xi+\gamma)/e2$ and
$\beta=\alpha\nu-\eta$, with the same assumption that in the latter
case. The inset of Fig. 2 and Fig. 4 show all the distributions
for the simulations when the finite size scalings were applied.
The best collapses were reached with the critical exponents shown
in Table 1. The match between $\nu$ values guarantees the quality
of the fits. A simple inspection of the $\beta$ values from the
simulation indicates that the correct assumption is constant
$\beta$. Until here, we have shown that, within the BTW model, the
distributions of avalanches measured in proportional portions of
the grid behave similarly as in the whole system; i.e., same
values of $\alpha$, and $\nu$ very close to $d$, with some
corrections which depend on how far from the center of the system
they are. {\it However, the distribution of} off the edge {\it
avalanches do not show a power law behavior but collapse when
scaling relations are applied.} The critical exponents do not
correspond to their analogous from the internal avalanche
distributions, since their avalanches do not involve sites located
at constant fraction of the system size, but depend on the system
size as $F=1/L$. The relation $\beta/\nu=2$ is a consequence of  $\langle s \rangle=1$
 for the {\it off the edge} avalanches, and it is independent from the system dimension.
 Now we present experiments in real piles
where these ideas are corroborated.

\begin{figure}[t]
\includegraphics[height=2.21in, width=2.8in]{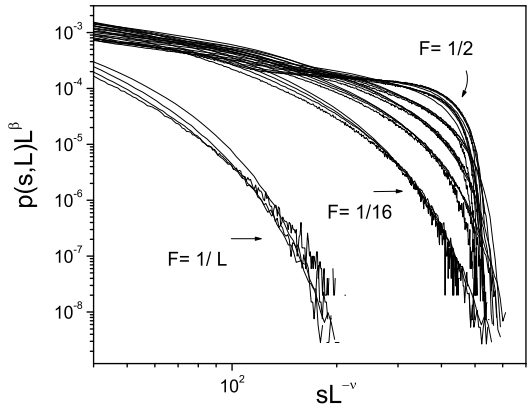}
\vspace{-0.5cm} \caption{\label{fig:f4} Scalings based on  the
ansatz $p(s,L)L^{\beta}=f(sL^{-\nu})$ for the curves of Fig. 3.
The critical exponents are reported in Table 1.}
\vspace{-0.5cm}
\end{figure}
The experimental setup is similar to our previous one \cite{
Altshuler et al-2001} and it consists of an acrylic strip
sandwiched between two parallel vertical glass plates $5.0 \pm
0.2\, mm$ apart from each other so that a horizontal surface of $5
\times L\, mm^2$ (where $L=24~cm$, $32~cm$) was available for the
formation of a pile of $4.000 \pm 0.005\, mm$ diameter steel
beads. The base consists of a row of the same type of beads glued
to the surface with random spacings (0, 1, 2 or 3 mm) between
them.
\begin{figure}[t]
\includegraphics[height=2.48in, width=3.3in]{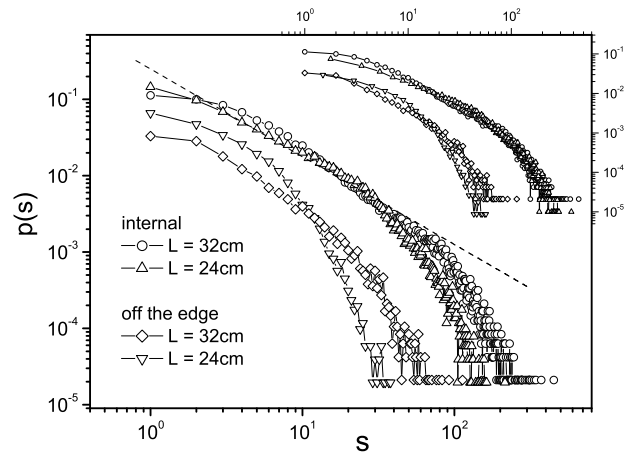}
\vspace{-0.5cm} \caption{\label{fig:f5} Avalanche size
distributions (normalized to the total number of avalanches) for
the experiment. The dashed line has a slope equal to -1.15. The
scalings based in  the ansatz $p(s,L)L^{\beta}=f(sL^{-\nu})$ are
shown in the inset. The critical exponents appear in Table 1. }
\vspace{-0.5cm}
\end{figure}
\begin{figure}[b]
\vspace{-1.5cm}
\includegraphics[height=2.96in, width=3.4in]{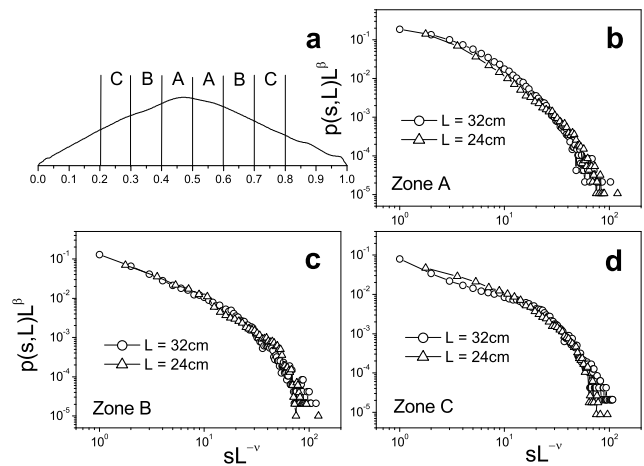}
\vspace{-0.5cm} \caption{\label{f6}(a) Schematic representation of
the different zones where
  internal avalanches were measured. The profile is the average of two
  system length profiles (normalized to the system length). (b, c, d) Scalings based in  the ansatz
$p(s,L)L^{\beta}=f(sL^{-\nu})$ for the internal avalanche
distributions for
  two system length in the different zones. The
critical exponents appear in Table 1.}
\end{figure}
\begin{table*}[t!]
\caption{Critical exponents for avalanche size distributions. The
experimental analogous of $F$ sites are zones A, B, C (see Fig.
6a).}\label{tab01}
\begin{tabular} {c c c c c c c c } \hline \hline
   avalanche&~~analytic&  &~~~~~~~~~~~~~~~~~~~&~simulations&~~~~~~~~  &~~~experiment&  \\
   distribution& $\nu$ & $\beta(\alpha=1)$ & $\alpha(\beta=1.45)$ & $\nu$ & $\beta$  &$\nu$ & $\beta$ \\ \hline
  internal & $~~2~~$ & $~~2~~$ & - & $2.00\pm 0.02$ & $2.20\pm 0.05$
    &~~~~$2.0\pm 0.1$ & $2.3\pm 0.1$ \\
 ~~ off the edge~~ &~ $0.67\pm 0.02$~ &~ $1.04\pm 0.03$~ &~ $1.61\pm 0.07$~ &~ $0.68\pm 0.02$~
 &$1.40\pm 0.05$ &~~~~$1.4\pm 0.1$~ &~ $2.8\pm 0.1$~ \\ \hline
  $~F=1/16$ & $1.16\pm 0.01$ & $1.53\pm 0.02$ & $0.93\pm 0.03$ & $1.18\pm 0.02$ & $1.45\pm 0.05$ &  &  \\
  $F=1/8$ & $1.13\pm 0.01$ & $1.51\pm 0.02$ & $0.96\pm 0.03$ & $1.11\pm 0.02$ & $1.45\pm 0.05$
   &~~(C)~ $2.0\pm 0.1$  & $2.8\pm 0.1$ \\
  $~F=3/16$ & $1.10\pm 0.01$ & $1.47\pm 0.02$ & $0.98\pm 0.03$ & $1.07\pm 0.02$ & $1.35\pm 0.05$ &  &
\\
  $F=1/4$ & $1.08\pm 0.01$ & $1.45\pm 0.02$ & $1.00\pm 0.03$ & $1.06\pm 0.02$ & $1.40\pm 0.05$
   &~~(B)~  $2.0\pm 0.1$  & $2.3\pm 0.1$ \\
  $~F=6/16$ & $1.05\pm 0.01$ & $1.42\pm 0.02$ & $1.03\pm 0.03$ & $1.04\pm 0.02$ & $1.35\pm 0.05$ &  &  \\
  $F=1/2$ & $1.04\pm 0.01$ & $1.41\pm 0.02$ & $1.04\pm 0.03$ & $1.04\pm 0.02$ & $1.40\pm 0.05$
   &~~(A)~  $2.0\pm 0.1$ & $ 2.1\pm 0.1$ \\  \hline \hline
\end{tabular}
\end{table*}
They were dropped one by one from a height of $10\pm 2\,mm$ above
the apex of the pile. Both lateral sides of the base were open, so
the beads were able to fall off the system. The whole setup was
computer controlled in such a way that a digital camera took an
image of the pile and then a new bead was added only after all
motion associated to the previous impact had stopped. A parallel
software found the center of each bead, and stored its position.
We define an internal avalanche of size $s$ when $s$ beads moved
in the whole pile (we define ``movement" as a displacement of the
center of a bead not smaller than $1/4$ of its diameter when
successive images are compared). The experimental equivalent of
the $F$ sites studied in the simulations, was the measurement of
avalanches (movements) in different portions of the pile (see Fig.
6a). An {\it off the edge} avalanche has a size $s$ when $s$ beads
fell off the pile after a dropping event. Each experiment included
more than 50~000 avalanches, with an average total duration of
250~h. The avalanches previous to the attainment of the ``steady"
average pile size were not taken into account for the statistics.

Fig. 5 displays the size distribution of the internal avalanches
and of {\it off the edge} ones for two system lengths. The
distributions of internal avalanches show a power law regime over
two decades with a slope equal to -1.15. The critical exponents
(also in Table 1) are $\nu=2$ and $\beta=2.3$, very close to those
expected from the BTW model in two dimensions. This demonstrates
that our system is not truly one-dimensional, but that the
avalanches penetrate the bulk of the pile. The distributions of
{\it off the edges} avalanches, in agreement with the previous
simulations, display curves that are not power laws but collapse
when appropriate scaling relations are applied. In our case it
happened with $\nu=1.4$ and $\beta=2.8$. The same exponents were
found when the {\it gapran} curves from \cite{ Altshuler et
al-2001} were re-normalized to the total number of avalanches and
the scaling relations applied. A value of $\nu>1$ indicates that
the avalanches that leave the pile involve more than one
dimension, i.e., more than a layer of one bead width on the
profiles \cite{next}. The exponent values are associated with
geometrical properties of the system, but they are related each other by
$\beta/\nu=2$. From this view, the power laws
experimentally found in {\it off the edge} avalanches
distributions in conical piles may be due to their similitude to
the internal ones as a consequence of the small size of the
systems.

The avalanches in different portions of the pile (Fig. 6) behave
in a similar way to the internal ones for the whole system showing the same
values of $\nu$. The differences in area between the zones provoke variations
in the values of $\beta$. This was corroborated reducing the width of the zone
B in a 30$\%$. Then both its area and its $\beta$ value coincide with their 
analogous for the zone C. All this reaffirms the
fact that the differences between the critical exponents between
{\it off the edge} and internal avalanches are due to the fact
that the former ones involve zones that are not a fixed proportion
of the system length.

In conclusion, we have explained why several experiments involving
{\it off the edge} avalanches do not show the expected power laws
in their avalanche distributions, but they do follow scaling
relations. We have also demonstrated the existence of SOC behavior
in a pile of grains provided its components are able to establish
a high degree of inner disorder, needed to spread activity in the
totality of the system. As a consequence scalings of {\it off the
edge} avalanche size distributions can be safely taken as an
indication of power law behavior of the internal avalanche size
distribution in a model sandpile.

We thank C. Noda and O. Ar\'es for collaboration in the
experiments, G. Rojas-Lorenzo for his help at the computer cluster
at InSTEC and R. Mulet for valuable suggestions. The ACLS/SSRC
Working Group on Cuba is acknowledged for providing access to
several journals.

\end{document}